\documentstyle[aps]{revtex}
\newcommand{\BE}{\begin{equation}}
\newcommand{\EE}{\end{equation}}
\newcommand{\BA}{\begin{eqnarray}}
\newcommand{\EA}{\end{eqnarray}}

\begin{document}
\draft

\title{On localization of acoustic waves}
\author{Zhen Ye, Haoran Hsu, Emile Hoskinson, and Alberto Alvarez}
\address{Department of Physics and Center for Complex Systems,
National Central University, Chung-li, Taiwan, ROC}

\date{Nov. 14, 1998}
\maketitle

\begin{abstract}

The paper presents a discussion on localization of acoustic waves.
Some important questions about wave localization are addressed. Particular
attention is paid to acoustic localization in liquid media containing
many air-filled bubbles. It is shown that an amazing collective behavior
appears when localization occurs. The study sheds new light to the much
discussed subject.

\end{abstract}

\pacs{PACS numbers: 43.20., 71.55J, 03.40K. To appear in Chin. J. Phys. 1999}

\section{Introduction}

When propagating through media containing many scatterers, waves will be
scattered by each scatterer. The scattered waves will be scattered again by
other scatterers. This process is repeated to establish an infinite recursive
pattern of rescattering between scatterers, forming a multiple scattering
process which causes the scattering characteristics of the scatterers to
change. Multiple scattering of waves is responsible for a wide range of
fascinating phenomena. This includes, on large scales, twinkling light in
the evening sky, modulation of ambient sound at ocean surfaces\cite{Vagle},
acoustic scintillation from turbulent flows\cite{DMF} and fish
schools\cite{YT}. On smaller scales, phenomena such as white paint, random
laser\cite{Laser}, electron transport in impured solids\cite{Im} and
photonic band gaps in periodic structures\cite{Band} are also explained
in terms of multiple scattering. Under proper conditions, multiple scattering
leads to a phenomenon termed {\it localization}, which has now become an
everyday experience.

Wave localization is a ubiquitous phenomenon. It refers to situations in
which transmitted waves in a scattering medium are trapped in space and
will remain confined in the neighborhood of the initial site until
dissipated. The concept of localization was conceived from the theory
describing the disorder induced conductor-insulator transition in
electronic systems\cite{Anderson}. Since its inception, wave localization
has stimulated considerable interest among scientists from many areas of
disciplinary. Several monographs have been devoted to the subject
(e.~g. Ref.~\cite{Sheng1}). Wave localization may be realized in a variety
of situations. In disordered solids, electron localization is common.
Localization has also been reported for microwaves in random scattering
rods and spheres\cite{Microwaves1,Microwaves2}, and recently for light
in a ground gallium-arsenide suspension in methanol\cite{Light}.
Acoustic localization has also been studied both theoretically\cite{Theory}
and experimentally\cite{Exp}. Research suggests that acoustic localization
may be observed in bubbly liquids\cite{Theory,YA}.

Despite the tremendous efforts, however, no deeper insight into localization
has been documented in the literature. The general cognition remains that
enhanced backscattering is a precursor to wave localization and that disorder
is an essential ingredient of localization. Important questions such as how
and when the localization occurs stay as an unsolved puzzle.

In a recent paper\cite{YA}, we proposed to study wave localization
phenomenon by investigating wave propagation in liquid media containing
many air-filled bubbles. There are several advantages of studying sound in bubbly liquids. 
(1) The air-filled bubbles are strong acoustic scatterer. At low frequencies, about
$ka \sim 0.0136$, it appears the resonant scattering and the scattering
strength is greatly enhanced at resonance, making it an ideal system
to study strong scattering. Here $k$ is the acoustic
wavenumber in water and $a$ is the radius of the bubbles.
(2) The scattering function of a single bubble has been
well studied and has a simple form. The scattering
function of a spherical bubble can be found in many textbooks\cite{Morse},
whereas the scattering function of a deformed bubble,
such as an ellipsoidal bubble, has also been analytically derived
recently\cite{Ye2}.
It is shown that the scattering
function of a single bubble has a simple isotropic resonant form,
permitting many simplifications. Furthermore, perhaps
more important, such an isotropic scattering feature remains valid even as
the bubbles are subject to significant deformation\cite{Ye2}. 
(3) Each term in the scattering function has clear physical meaning,
and can be modified
according to the need. When the thermal exchange and viscosity effects are
taken into account, absorption indeed shows up but in a way such that
it can be turned off or adjusted. This allows to unequivocally
isolate localization due to scattering from attenuation caused by absorption,
making waves in bubbly water an ideal system
for studying phenomena related to multiple scattering.

In this paper, we first discuss general aspects of wave localization in a 3D
system. An intuitive picture is proposed to describe wave localization and
is subsequently supported by
numerical examples in acoustic propagation in bubbly liquids. Wave
localization is a phase transition.
A novel method is proposed to describe such a phase transition. It is shown
that when the localization occurs, all air-filled bubbles as a resonant
scatterer prevail a surprising collective behavior.

\section{General aspects}

Consider a plane wave normally
incident upon a semi-infinite random medium. The transport equation for
the total energy intensity $I$ may be intuitively written as
\BE
\frac{dI}{dx} = -\alpha I, \label{eq:1}
\EE
where $\alpha$ represents decay along the path traversed. After penetrating
into the random medium, the wave will be scattered by random
inhomogeneities. As a result, the wave coherence starts to decrease,
yielding the way to incoherence. Extinction of the coherent intensity
$I_C$ is described by
\BE
\frac{dI_C}{dx} = -\gamma I_C, \label{eq:2}
\EE
with the attenuation constant $\gamma$. Eqs.~(\ref{eq:1}) and
(\ref{eq:2}) lead to the exponential solutions
\BE
I(x) = I(0)e^{-\alpha x}, \ \ \ \mbox{and} \ \ \
I_C(x) = I(0)e^{-\gamma x}.
\EE
In deriving these equations, the boundary condition was used; it states
that $I(0) = I_C(0)$ as no scattering has been incurred yet at the interface.
According to energy conservation, the incoherent intensity $I_D$ (diffusive)
is thus
\BE
I_D(x) = I(x) - I_C(x).
\EE

When there is no absorption, the decay constant $\alpha$ is expected
to vanish and the total intensity will then be constant along the
propagation path. Then, the coherent energy gradually decreases due
to random scattering and transforms to the diffusive energy, while the
sum of the two forms of energy remains a constant.
This scenario, however, changes when localization occurs.
Even without absorption, the total intensity can be localized near the
interface due to multiple scattering. When this happens, $\alpha$ does
not vanish. The transport of the total intensity may be still described by
Eq.~(\ref{eq:1}), and the inverse of $\alpha$ would then refer to the
localization length.

The above perceptual description may be illustrated by Fig.~1.
Without or with little absorption, the energy propagation is anticipated
to follow the behavior depicted in (a). When the localization comes in sight,
the wave will be trapped within an $e$-folding distance from the penetration,
as prescribed by (b). In the non-localization case, the diffusive intensity
increases steadily as more and more scattering occurs,
complying with the Milne diffusion\cite{Morse}. In the
localized state, the diffusion energy increases initially.
It will be eventually stopped by the interference
of multiple scattering waves. Issues may be raised with respect to whether
this apprehended image is supported by actual situations. In the rest, we
will inspect this problem by considering acoustic waves in bubbly water.

\section{Acoustic localization in bubbly liquids}

Consider sound emission from a bubble cloud. For simplicity, the shape
of the cloud is taken as spherical. Such a model eliminates irrelevant
edge effects, and is useful to separate phenomena pertinent to the discussion.
Total $N$ bubbles of the same radius $a$ are randomly distributed
within the cloud. The volume void fraction, the space occupied by bubbles
per unit volume, is take as $\beta$. A monochromatic acoustic source is
located at the center of the cloud. Adaptation of such a model for other
geometries and situations is straightforward. The wave transmitted from
the source propagates through the bubble layer, where multiple scattering
incurs, and then it reaches a receiver located at some distance from the
cloud. The multiple scattering in the bubbly layer is described by
a set of self-consistent equations. The energy transmission
can be solved numerically in a rigorous fashion\cite{YA}.

A traditional way to study wave localization is to calculate the
Green's function for the energy transport, leading to the Bethe-Salpeter
equation. Under certain approximations, such as long time and
large distance from the initial site, a solution to this equation can be
obtained in the form of a diffusion formula in which a mean free path and
diffusion coefficient can be defined and have been used as the basis
for discussion in the literature. Alternatively, certain situations allow
direct computation of the energy transport, without recourse to unnecessary
approximations. In the case, information about wave propagation can be
inferred straightway. Acoustic wave propagation is one of such situations.
The propagation is calculated incorporating all multiple scattering
by the self-consistent scheme\cite{YA}. In the following, we will inspect
the features of acoustic localization from three aspects and then
present a discussion.

\subsection{Frequency response}

First consider the frequency response. In Fig.~2, the transmission
in arbitrary units is plotted against frequency in terms of $ka$ for two
different bubble sizes. Here $k$ is the usual wavenumber, and the bubble
void fraction is $10^{-3}$. It is clearly suggested in the figure that a
narrow region is opened in which the transmission is virtually forbidden. This
inhibition gap ranges roughly from $ka = 0.015$ to $0.02$. 
When the void fraction is reduced to about $10^{-5}$, the
localization disappears\cite{AY}.
In order to show that the inhibition is not
because of absorption, the situation with the absorption being turned off
is also plotted in the dotted line for $a = 2$cm. The comparison between
with and without absorption excludes the absorption as the cause for the
propagation hindrance. In fact, when the absorption factor
of a single bubble is increased, the localization will be degraded.

\subsection{Distance variation}

To unambiguously identify that the transmission inhibition region
is indeed the localization range, it is proper to study the spatial dependence of energy propagation.
Fig.~3(a) presents the total transmission and its coherent
and diffusive parts, scaled by the geometry
spreading factor $r^2$, as a function of propagation distance scaled by the
radius of the bubble cloud. The solid, dotted and broken
lines refer to total, diffusive and coherent intensity
respectively. The bubble radius is 2 cm, whereas
the frequency is taken as $ka = 0.171$, which lies in the localization
regime.
When plotting the data in the natural logarithm in Fig.~3(b), we
found that nearly all the data rest on a straight line. In (b), the straight
line refers to the fitted curve and the crosses refer to the numerical data.
We find the slope for the line amounts to a constant
$\alpha = -0.195$, which is found to be true for other bubble sizes as well.
This suggests that the intensity varies with propagation distance $r$ as
\BE
I \sim  \frac{e^{-\alpha r/a}}{r^2}. \label{eq:5}
\EE
Therefore the transport equation for the total wave can be written as
\BE
\frac{dI}{dr} = -\left(\frac{\alpha}{a} + \frac{2}{r}\right)I,
\label{eq:6}
\EE
which is an equivalence of Eq.~(\ref{eq:1}) in the spherical geometry.
The second term on the right hand side of Eq.~(\ref{eq:6}) denotes the
geometrical spreading
effect. From Eq.~(\ref{eq:5}), the localization length is computed to be
$l = 5.13a$. At this length, $kl \approx 0.0877$; therefore the
Ioffe-Regel criterion for localization\cite{Sheng1} is satisfied, providing
another verification. At this frequency we also calculated attenuation
length due to thermal and viscosity absorption to be about $118a$, which is
much larger than the observed localization length.

The same calculation has also been performed for other sizes of bubbles.
We found that within the localization regime, the ratio between the
localization length and the size of bubble is nearly a constant around $5$,
as long as the bubble radius is large enough, roughly bigger than 10$\mu$m.
For too small bubble, the absorption due to thermal exchange and viscosity
effects is significant. Then the transmission will be size dependent and
therefore the scaling behavior disappears.
These features do not appear for $ka$ outside the localization regime.
Fig.~3(a) draws remarkable analogy to the perception demonstrated
in Fig.~1(b).

\subsection{Collective behavior}

Upon incidence, each air-bubble acts as a secondary pulsating point
source. The
radiated wave from the $i$-th bubble ($i = 1, 2, 3,..., N$)
can be written as
$A_iG_0(\vec{r}-\vec{r}_i)$, where $G_0(\vec{r}-\vec{r}_i)$
is the usual 3D Green's function and $\vec{r}_i$ denotes the positions of
the bubble. The complex coefficient $A_i$ refers to the effective
strength of the secondary source,
and is computed incorporating all multiple scattering effects. The total wave
at any space point is the addition of the direct wave from the transmitting
source and the radiated wave from all bubbles.

We express $A_i$ as $|A_i|\exp(i\theta_i)$; the modulus $A_i$
represents the strength, whereas $\theta_i$ the phase of the secondary
source. We assign a unit vector $\vec{u}_i$, hereafter termed
phase vector, to each phase $\theta_i$, and these vectors are
represented on an argand diagram in the $x-y$ plane. That is, the
starting point of each phase vector is positioned at the center of
individual scatterers with an angle with respect to the positive
$x$-axis equal to the phase, $\vec{u}_i = \cos\theta_i\hat{x} +
\sin\theta_i\hat{y}$.
Letting the phase of the initiative emitting source be zero,
numerical experiments are carried out to study the behavior of the phases
of the bubbles and the spatial distribution of the acoustic energy.
Figure~4 shows the argand diagrams of
the phase vectors, and the energy distribution for three frequencies in
terms of $ka$ for one arbitrary
random configuration of bubbles.

We observe that for frequencies smaller than a certain value, there is no
obvious order for the directions of the phase vectors, nor for the energy
distribution. The phase vectors point to various
directions, and no energy localization appears.
The random behavior in the directions and energy
distribution is attributed to the boundary effect of a finite number of the
scatterers. In effect, as the wave is not localized, it can propagate through
and is reflected by the asymmetric border; all bubbles
can experience the effect via strong multiple scattering.

As the frequency increases, an order in the phase vectors and energy
localization becomes evident. The case with $ka = 0.0164$
clearly shows that the energy
is localized near the source, and amazingly, all bubbles oscillate
completely in phase, but exactly out of phase with the transmitting source.
Such collective behavior allows for efficient cancellation of incoming
waves. 
The energy distribution decays exponentially, which sets
the localization length. The localization behavior is independent of
the outer boundary and always appears for sufficiently large $\beta$ and $N$.
When the frequency increases further, exceeding a certain amount, the
in-phase order disappears. Meanwhile, the wave becomes non-localized again.
This is illustrated by the case of $ka = 0.1$.

Further numerical investigation shows that the pattern depicted by
Fig.~4 holds qualitatively true for other sizes of bubble (as long as the
bubble is not too small; when the bubble is too small,
the viscosity and thermal effects will dominate and the localization
phenomenon will disappear. And the features are always
valid for sufficiently large bubble void fraction.

\subsection{On localization}

Now a question immediately follows: What can we learn about wave localization
from the above discussion? First, wave localization
refers to trapping of the total wave energy, which may be
indicated by the exponential decay of transmitted energies along the
distance traveled by wave. However, the localization due to the scattering
must be differentiated from the attenuation due to absorption. It is
well-known that when absorption is present, energies will also decrease with
the traveling distance. Second, when localization occurs, wave is trapped
near the point of transmission. If a continuous wave is pumped into the
system, there could be several conjectures. (1) The energy will be built up
in the neighborhood of the transmission point until the amplitude is so large
that localization fails. (2) When the localization occurs, no more energy can
be pumped into the system. This scenario may be hinted by from Fig.~4
which shows that the anti-phase collective behavior allows for efficient
cancellation of transmitting waves. This may be in analogy with situation for
the electrical current in a conductor with resistance $R$ that is connected
to a battery with potential $V$. The electrical power is $V^2/R$. When added
with sufficient amount of impurities, the conductor will become an insulator,
i.~e. $R\rightarrow \infty$. Then the battery can no longer inject power
into the medium. Which scenario describes the actual situation remains an
open problem. However, the present study seems to hint at the second
hypothesis.

\section{Summary}

In summary, we have considered the behavior of acoustic localization
in water containing many air-filled bubbles, using a simple numerical
model. The localization behavior was investigated and is shown to follow
the description of a simple transport equation.
The research provides a fundamental backbone for a simple intuitive picture
about wave localization in random media. A novel approach has been proposed
for describing the localization phase transition. It is shown that
a collective behavior appears when wave localization occurs. More detailed
discussion of the collective behavior and a consideration of 2D situations
will be published elsewhere\cite{YH}.

\section*{Acknowledgments}

The work received support from National Science Council of ROC
and also from the National Central University in the
form of a CO-OP scholarship to EH. One of us (AA) also acknowledges the
support from NSC and the Spanish Ministry of Education in the form of
post doctoral fellowships.

\section*{Figure Captions}

\begin{description}
\item[Fig. 1] A conceptual illustration of wave localization
\item[Fig. 2] Transmission as a function of $ka$ for two different bubble radii
\item[Fig. 3] Transmission as a function of propagation distance and localization length
\item[Fig. 4] Left column: Argand diagrams for the two-dimensional phase
vectors lying parallel to the x-y plane. Right column: Spatial distribution of acoustic energy (arbitrary scale).
\end{description}

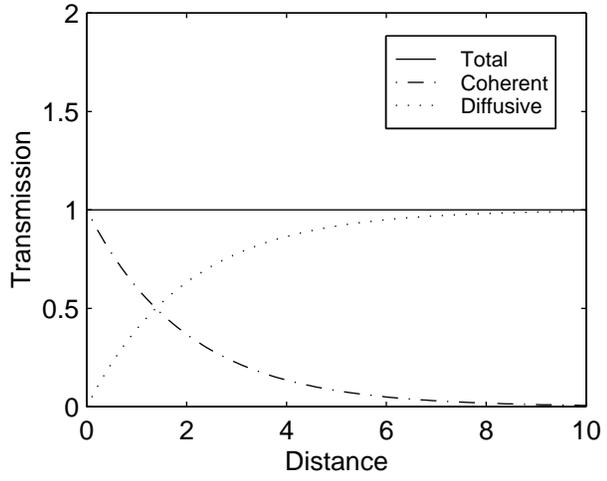
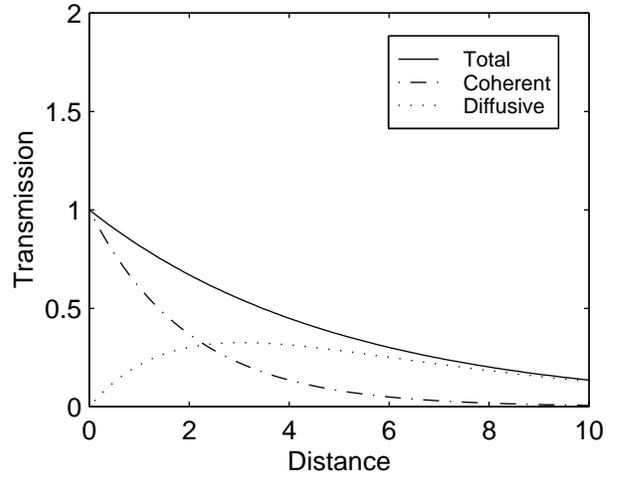
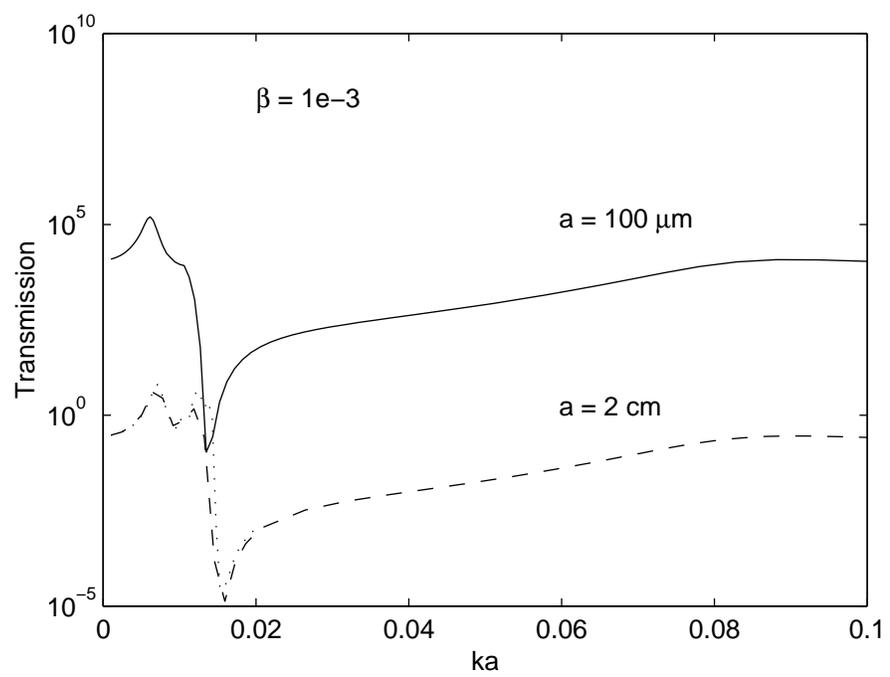
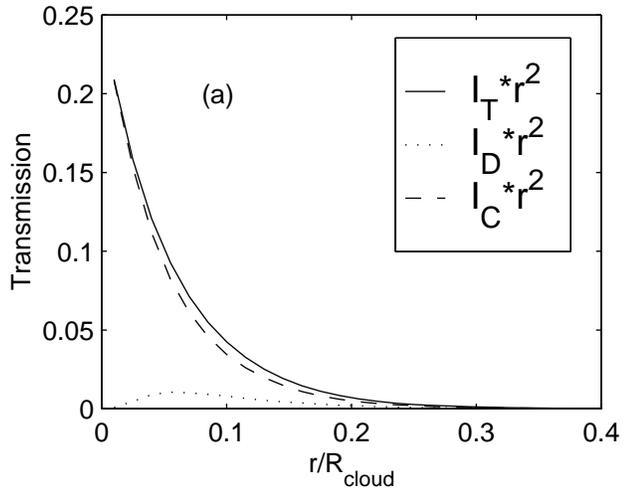
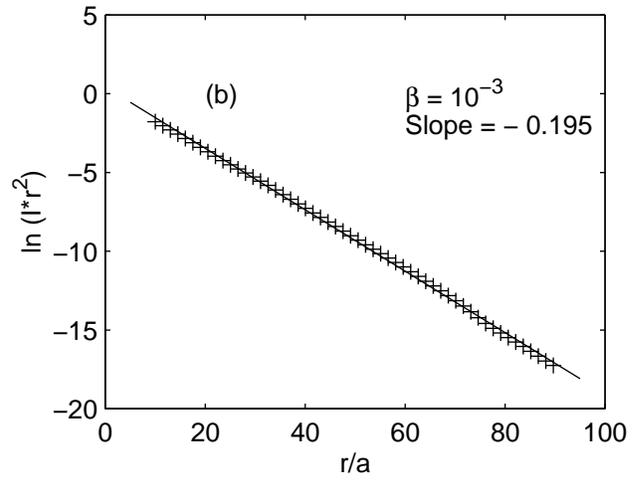
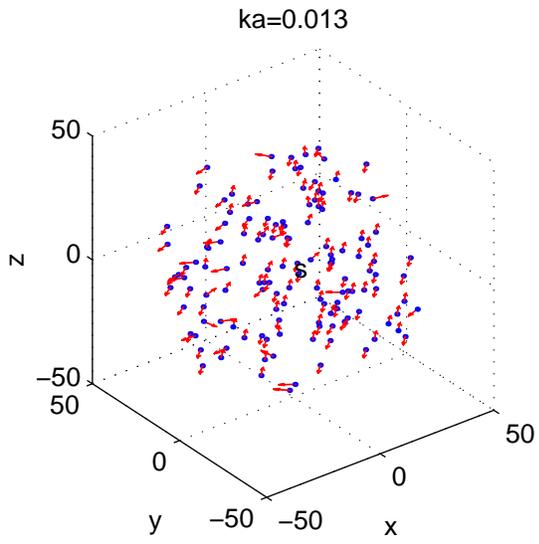
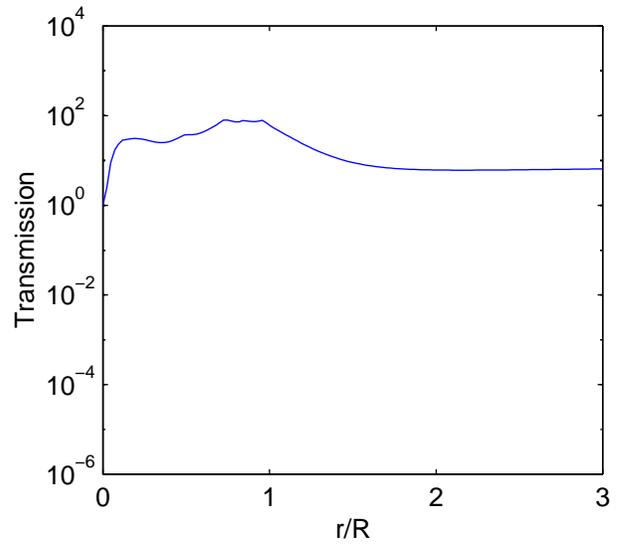
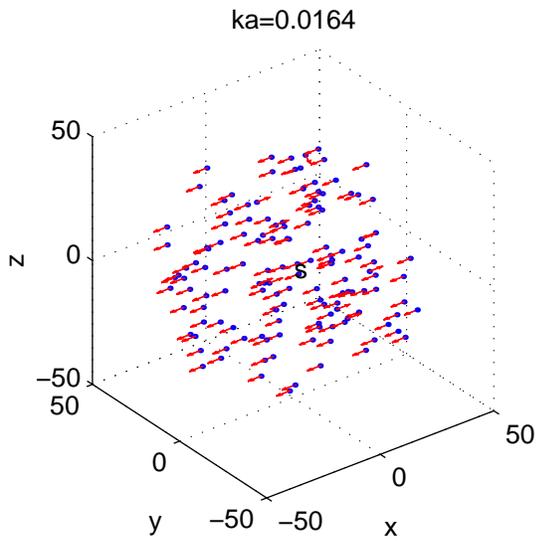
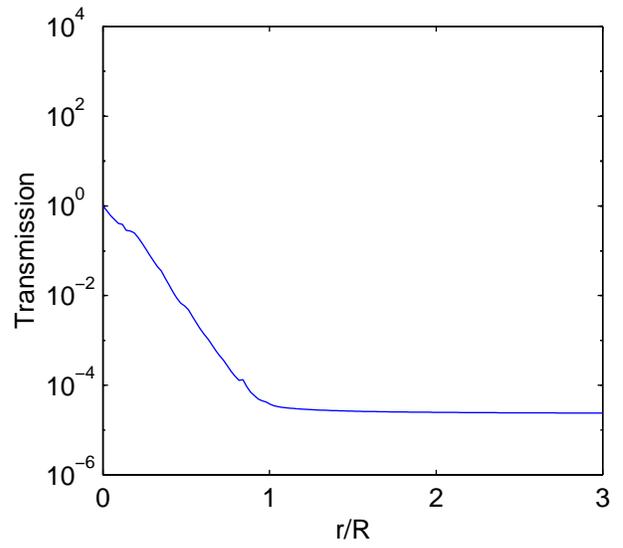
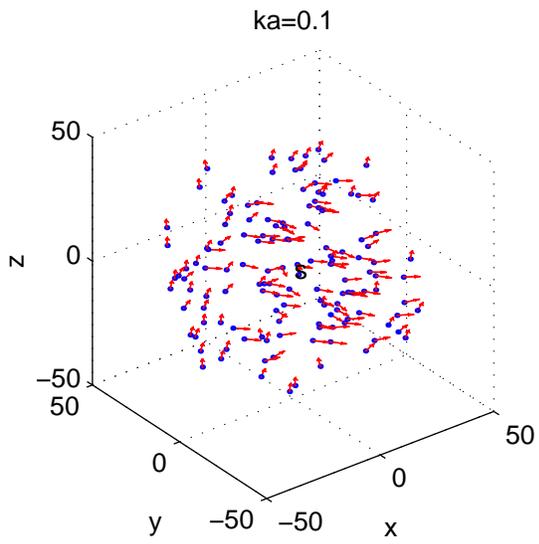
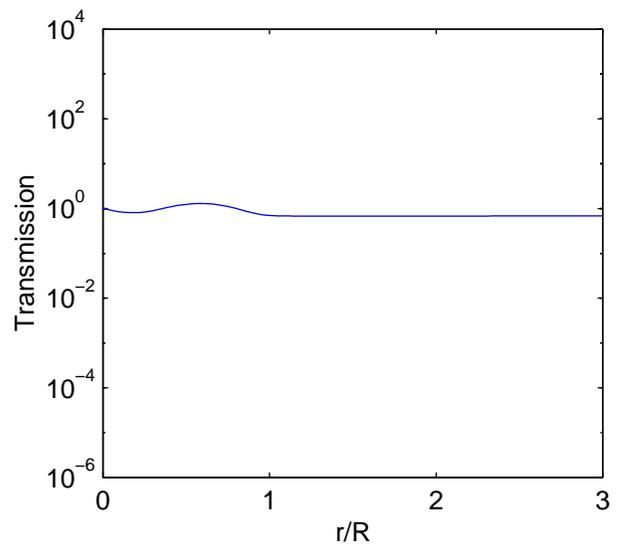
\end{document}